\documentclass[prb,showpacs]{revtex4}

\usepackage{amsmath, amsthm}

\newcommand{\tint}{{\textstyle \int}}

\begin{document}

\title{Quantum mechanics using Fradkin's representation} 

\author{K. V. Shajesh}
\email{shajesh@nhn.ou.edu}
\homepage{http://www.nhn.ou.edu/~shajesh}
\author{Kimball A. Milton} 
\email{milton@nhn.ou.edu}
\homepage{http://www.nhn.ou.edu/~milton}
\altaffiliation{On sabbatical at Department of Physics,
Washington University, St. Louis, MO 63130 USA}
\affiliation{Oklahoma Center for High Energy Physics 
and Homer L. Dodge Department of Physics and Astronomy,
University of Oklahoma, Norman, OK 73019 USA}

\date{\today}
\pacs{03.65.Db, 11.10.Kk, 11.15.Tk} 

\begin{abstract}
Fradkin's representation is a general method of attacking problems
in quantum field theory, having as its basis the functional approach of
Schwinger.  As a pedagogical illustration of that method, we explicitly
formulate it for quantum mechanics (field theory in one dimension) and 
apply it to the solution of Schr\"odinger's equation for the quantum 
harmonic oscillator. 
\end{abstract}

\maketitle


\section{Introduction}

In one of his papers\cite{fradkin:1966}
on functional methods
E. S. Fradkin derives a formal solution
(Fradkin's representation\cite{fried}) to a class of
differential equations which occur in quantum field theory.
The solution of the problem reduces 
to operations involving functional derivatives alone.
As such, it is complementary to the far more familiar solution
in terms of Feynman functional integrals.
In a different context, Fradkin mentions that for
the case of a harmonic oscillator
``all operations are quite simple and brought
to completion.''
In this pedagogical note  we carry out this ``simple'' task
(which is perhaps not quite so straightforward as it might appear),
as an example of the use of this method, which is not widely
discussed.  Elsewhere, we will apply this method to nontrivial physical
examples, particularly to describe the physics of magnetic monopoles.

Fradkin's representation was developed in the 
context of quantum field theories and in this
regard it is an extension to
Schwinger's functional representation\cite{sch-z} for 
the vacuum to vacuum persistence probability amplitude 
$Z \equiv \langle 0_+ | 0_- \rangle$.
The latter object, expressed in terms of external sources,
generates all the Green's functions of the theory, and therefore
defines the theory.  Schwinger's expression of field theory in this way
was the content of important papers published in the 1950s.\cite{pnas}


\section{Statement of the problem}
Schr\"odinger's equation for a quantum mechanical oscillator 
in one dimension is
\begin{equation}
i \hbar \frac{\partial}{\partial t} \psi (x,t)
= \left[ - \frac{\hbar^2}{2 m} 
           \frac{\partial^2}{\partial x^2}
         + \frac{1}{2} k x^2
  \right] \psi (x,t).
\label{sch_eqn}
\end{equation}
In terms of the parameters 
$\omega$ and $\alpha$, defined as
$\frac{k}{\hbar} = \alpha^2 \omega$
and $\frac{m}{\hbar} = \frac{\alpha^2}{\omega}$,
we can construct the dimensionless pairs of variables,
$\omega t$ and $\alpha x$. Without any loss of 
generality, we can choose $\omega = 1$ and $\alpha = 1$,
which renders $t$ and $x$ dimensionless.
In terms of dimensionless $t$ and $x$ Schr\"odinger's 
equation takes the form
\begin{equation}
i \frac{\partial}{\partial t} \psi (x,t)
= \left[- \frac{1}{2} \frac{\partial^2}{\partial x^2}
         + \frac{1}{2} x^2 \right] \psi (x,t).
\label{sch_eqn1}
\end{equation}
The solution to this differential equation
for a particular initial condition 
\begin{equation}
\psi (x,0) = \pi^{-\frac{1}{4}}
e^{- \frac{1}{2} (x - a)^2}
\label{in_con}
\end{equation}
is\cite{schiff:qm} 
\begin{equation}
\psi (x,t) = \pi^{-\frac{1}{4}}
 e^{- \frac{1}{2} (x - a \cos t)^2}
 e^{-\frac{i}{2}[t + 2 a x \sin t 
       - \frac{1}{2} a^2 \sin 2 t]},
\label{soughtresult}
\end{equation}
where $a$ is interpreted as the 
amplitude of the corresponding classical oscillator.
The goal here is to reproduce this result using Fradkin's 
representation for Schr\"odinger's equation.


\section{Fradkin's representation for Schr\"odinger's equation}

In this section we shall keep our treatment 
general for a time-dependent potential of the form $U(x,t)$.
Consider the following differential equation,
constructed by introducing a suitable source term $v(t)$ for the
derivative operator in Schr\"odinger's equation, 
\begin{equation}
\frac{\partial}{\partial t} W [x,t;v]
= \left[ \frac{i }{2} \frac{\partial^2}{\partial x^2}
         + v(t) \frac{\partial}{\partial x}
         - i U(x,t) 
  \right] W[x,t;v],
\label{frad_sch_eqn}
\end{equation}
where $W[x,t;v]$ is a functional of the source $v$ 
and a function of $x$ and $t$. 
The solution $\psi (x,t)$ to the 
original Schr\"odinger equation is related
to $W[x,t;v]$ by the prescription
\begin{equation}
\psi (x,t) = \left\{ W[x,t;v] \right\}_{v=0}.
\label{psi_w}
\end{equation}
The initial condition to Eq.~(\ref{frad_sch_eqn}) 
is prescribed, in terms of the initial condition to the
original Schr\"odinger equation, to be 
\begin{equation}
W[x,0;v] = \psi (x,0). 
\end{equation}
Notice that we have chosen our initial condition to be 
independent of $v$, thus we have
\begin{equation}
\frac{\delta}{\delta v(\tau)} W[x,0;v] = 0.
\label{dvw0}
\end{equation}

Before going ahead, it is good to appreciate the 
unitary property of the differential equation (\ref{frad_sch_eqn}).
We know that Schr\"odinger's equation preserves the 
norm of $\psi (x,t)$,
\begin{equation}
\frac{d}{d t} (\psi, \psi)
= \frac{d}{d t} 
{\int} dx \left| \psi (x, t) \right|^2 
= 0,
\end{equation} 
which is the statement of probability conservation.
In other words, the time evolution of $\psi (x,t)$
is generated by a unitary transformation.
Using the traditional technique, which involves
multiplying Eq.~(\ref{frad_sch_eqn}) by the complex 
conjugate of $W[x,t;v]$, and then adding the modified equation 
to its complex conjugated version, we can see that 
\begin{equation}
\frac{d}{d t} 
{\int} dx~ \Big| W[x,t;v] \Big|^2 
= 0,
\end{equation}
if $v(t) = v(t)^*$. (We are also assuming that the potential $U$ is real.)
 This exposes a very welcome feature 
of Eq.~(\ref{frad_sch_eqn}), the time 
evolution involved in $W[x,t;v]$ is unitary if $v(t)$ is real.

To determine Fradkin's representation for 
Schr\"odinger's equation we start by integrating
Eq.~(\ref{frad_sch_eqn}) with respect to time from 
0 to $t$ and obtain the following integral equation
\begin{equation}
W[x,t;v] = W[x,0;v]
 + \int_0^t dt^\prime
   \left[ \frac{i }{2 } \frac{\partial^2}{\partial x^2}
         + v(t^\prime) \frac{\partial}{\partial x}
         - i U(x,t^\prime) 
   \right] W[x,t^\prime;v]. 
\label{int_w}
\end{equation}
Using the above integral equation 
in conjunction with Eq.~(\ref{dvw0})
it is easy to derive a very useful identity 
\begin{equation}
\lim_{\epsilon \rightarrow 0}
\frac{\delta}{\delta v(t - \epsilon)}
W[x,t;v]
= \frac{\partial}{\partial x} W[x,t;v]
\label{dw=dxw},
\end{equation}
which involves the functional derivative of $W[x,t;v]$, and
$\epsilon$ is a positive infinitesimal quantity.  This,
of course, is the reason the source $v(t)$ was introduced.
The above identity will play a crucial role in  
the following development.
For the sake of dimensional analysis,
it might be helpful to remind ourselves that
the dimension of a functional derivative is not equal to the 
inverse of the dimension of the function,
because the functional derivative is defined as 
\begin{equation}
\frac{\delta f(x)}{\delta f(y)} = \delta (x - y),
\label{dfdf=d}
\end{equation}
where the delta function has the dimension 
of the inverse of the variable $x$.

\subsection{Proof of Eq.~(\ref{dw=dxw})}
We begin the proof of Eq.~(\ref{dw=dxw}) 
by introducing a variable $\tau$ which 
is defined in the domain $t - \epsilon < \tau < t$,
where $\epsilon$ is a positive infinitesimal quantity.
Starting from Eq.~(\ref{int_w}) and using Eq.~(\ref{dvw0})
we can write
\begin{eqnarray} 
\frac{\delta }{\delta v(\tau)} W[x,t;v] 
&=&
\int_0^t dt^\prime \delta (t^\prime - \tau)
\frac{\partial}{\partial x} W[x,t^\prime;v]
+ \int_0^t dt^\prime
   \left[ \frac{i }{2 } \frac{\partial^2}{\partial x^2}
         + v(t^\prime) \frac{\partial}{\partial x}
         - i U(x,t^\prime) 
   \right] 
\frac{\delta }{\delta v(\tau)} W[x,t^\prime;v],
\end{eqnarray}
where we have used the definition of functional derivative 
in Eq.~(\ref{dfdf=d}).
The last expression when iterated gives us
\begin{eqnarray}
\frac{\delta }{\delta v(\tau)} W[x,t;v]
&=&
\int_0^t dt^\prime \delta (t^\prime - \tau)
\frac{\partial}{\partial x} W[x,t^\prime;v]
\nonumber
\\
&& \mbox{}+ \int_0^t dt^\prime 
   \left[ \frac{i }{2 } \frac{\partial^2}{\partial x^2}
         + v(t^\prime) \frac{\partial}{\partial x}
         -i U(x,t^\prime)  
   \right] 
\int_0^{t^\prime} dt^{\prime \prime} \delta (t^{\prime \prime} - \tau)
\frac{\partial}{\partial x} W[x,t^{\prime \prime};v]
+ \cdots,
\label{dwdv_series}
\end{eqnarray}
which is a series expression for the functional derivative
of $W$ with respect to $v$.
The first term in the series trivially evaluates to our
desired result 
because $0<\{t^\prime,\tau\}<t$. In evaluating the second term 
we should be careful because the contribution from the 
delta function depends on whether or not $\tau$ is greater than
$t^\prime$. We evaluate the second
term by splitting the integral over $t^\prime$ at $t'=t-\epsilon$.
It is clear that only the integral over $t'$ from $t-\epsilon$ to
$t$ can contribute, and, by the mean value theorem, this integral
is of order $\epsilon$.  Higher terms in the series are even smaller,
so we conclude that in the limit $\epsilon\to0$ 
the result (\ref{dw=dxw}) holds true.

\subsection{Expression for wavefunction}
To proceed let us write the solution to $W[x,t;v]$ in 
terms of an auxiliary functional $W_1$:
\begin{equation}
W[x,t;v] = 
\lim_{\epsilon \rightarrow 0}
 e^{  \frac{i}{2} \int_0^t d \tau
     \frac{\delta^2}{\delta v(\tau - \epsilon)^2}}
 W_1[x,t;v],
\label{w=w1}
\end{equation}
which when substituted into Eq.~(\ref{frad_sch_eqn}) 
and after the use of Eq.~(\ref{dw=dxw}) gives 
\begin{eqnarray}
\frac{\partial}{\partial t} W_1[x,t;v] &=& 
\lim_{\epsilon \rightarrow 0}
 e^{ - {\frac{i}{2} \int_0^t d \tau
     \frac{\delta^2}{\delta v(\tau - \epsilon)^2}}}
 \left[v(t) \frac{\partial}{\partial x} 
       - i U(x,t)\right]
 e^{  \frac{i}{2} \int_0^t d \tau
     \frac{\delta^2}{\delta v(\tau - \epsilon)^2}}
 ~W_1[x,t;v],
\label{w1_eqn1}
\end{eqnarray}
the functional operator having removed the second derivative
term in Schr\"odinger's equation.
The exponential of the functional operator in the 
above equation contributes nothing when it acts on 
the quantity in the brackets due to the fact
that the sources are defined at different times, that is,  
because $\tau - \epsilon < t$.
As a consequence, the exponential terms in Eq.~(\ref{w1_eqn1})
nullify each other by passing through the bracketed quantity.
The differential equation for $W_1$ thus takes the form 
\begin{eqnarray}
\frac{\partial}{\partial t} W_1[x,t;v] 
=
\left[v(t) \frac{\partial}{\partial x} 
       - i U(x,t)\right]
W_1[x,t;v].
\label{w1_eqn}
\end{eqnarray}
Let us eliminate the 
$\frac{\partial}{\partial x}$ 
term by writing 
\begin{equation}
W_1[x,t;v] = 
~e^{  \int_0^t d \tau v(\tau) 
    \frac{\partial}{\partial x}} ~W_2[x,t;v],
\label{w1=w2}
\end{equation}
which when substituted in Eq.~(\ref{w1_eqn}) gives
\begin{eqnarray}
i \frac{\partial}{\partial t} W_2[x,t;v]
&=& e^{  - \int_0^t d \tau v(\tau) 
    \frac{\partial}{\partial x}}
 U(x,t)e^{  \int_0^t d \tau v(\tau) 
    \frac{\partial}{\partial x}}
  ~W_2[x,t;v]\nonumber
\\
&=& U\left( x - {\textstyle \int_0^t d \tau\, v(\tau)},t \right)W_2[x,t;v].
\label{w2_eqn}
\end{eqnarray}
The last equation can be immediately exponentiated,
\begin{eqnarray}
W_2[x,t;v] = W_2[x,0;v] 
   ~e^{  -i 
      \int_0^t dt^\prime
      U\left( x - {\int_0^{t^\prime} d \tau\, v(\tau)},t^\prime \right)}.
\label{w2=u}
\end{eqnarray}

Thus, following the sequential substitutions in  
Eqs.~(\ref{w=w1}), (\ref{w1=w2}), and (\ref{w2=u}),
we have a formal solution to the differential  
equation in Eq.~(\ref{frad_sch_eqn}): 
\begin{eqnarray}
W[x,t;v] = 
\lim_{\epsilon \rightarrow 0}
 e^{  {\frac{i}{2} \int_0^t d \tau
     \frac{\delta^2}{\delta v(\tau - \epsilon)^2}}}
 e^{  \int_0^t d \tau v(\tau) 
    \frac{\partial}{\partial x}} 
 W_2[x,0;v] 
   e^{  -i 
      \int_0^t dt^\prime
      U\left( x - {  \int_0^{t^\prime} d \tau v(\tau)},t^\prime \right)}.
\label{frad_soln}
\end{eqnarray}
Using the above formal representation for $W[x,t;v]$ 
we can write Fradkin's representation for  Schr\"odinger's equation
for the time-evolved wavefunction 
using the prescription in Eq.~(\ref{psi_w}) as
[because $W(x,0,v)=W_1(x,0,v)=W_2(x,0,v)=\psi(x,0)$] 
\begin{eqnarray}
\psi (x,t) &=& 
\left.
\lim_{\epsilon \rightarrow 0}
 e^{  {\frac{i}{2} \int_0^t d \tau
     \frac{\delta^2}{\delta v(\tau - \epsilon)^2}}}
 e^{  \int_0^t d \tau v(\tau) 
    \frac{\partial}{\partial x}} 
 \psi (x,0) 
   e^{  -i 
      \int_0^t dt^\prime
      U\left( x - {\int_0^{t^\prime} d \tau v(\tau)},t^\prime \right)} 
\right|_{v=0}\nonumber
\\
&=&
\left.
\lim_{\epsilon \rightarrow 0}
 e^{  {\frac{i}{2} \int_0^t d \tau
     \frac{\delta^2}{\delta v(\tau - \epsilon)^2}}}
 \psi \left(x + {\textstyle \int_0^{t} d \tau~ v(\tau)},0\right) 
   e^{  -i 
      \int_0^t dt^\prime
      U\left( x + {\int_{t^\prime}^{t} d \tau v(\tau)},t^\prime \right)} 
\right|_{v=0}.
\label{frad_sch_soln}
\end{eqnarray}


\section{Quantum harmonic oscillator}

For the case of the quantum harmonic oscillator
we have $U(x,t) = \frac{1}{2}  x^2$, 
for which Schr\"odinger's equation takes the form in 
Eq.~(\ref{sch_eqn1}), where we impose the initial condition
$\psi (x,0)$ given in Eq.~(\ref{in_con}).
Fradkin's representation for the quantum harmonic oscillator
is obtained by substituting
$U(x,t)$ and $\psi (x,0)$ into Eq.~(\ref{frad_sch_soln}),
which gives
\begin{eqnarray}
\psi (x,t) 
&=&
\left.
\lim_{\epsilon \rightarrow 0}
  \pi^{-\frac{1}{4}}
 e^{  \frac{i}{2}  
     \int_0^t d \tau \frac{\delta^2}{\delta v(\tau - \epsilon)^2}}
 e^{  - \frac{1}{2}  
  \left(x - a + \int_0^{t} d \tau v(\tau)\right)^2}
   e^{  -\frac{i}{2} 
      \int_0^t dt^\prime
      \left( x + {  \int_{t^\prime}^{t} d \tau v(\tau)} \right)^2} 
\right|_{v=0}.
\end{eqnarray}
For our purpose it will be convenient to rewrite the 
above equation in the general form
\begin{eqnarray}
\psi (x,t) 
=
\left.
\lim_{\epsilon \rightarrow 0}
  \pi^{-\frac{1}{4}}
 e^{  \frac{1}{2} 
     \int_0^t d t^\prime \int_0^t d t^{\prime \prime} 
     \frac{\delta}{\delta v(t^\prime - \epsilon)}
     A(t^\prime, t^{\prime \prime})
     \frac{\delta}{\delta v(t^{\prime \prime} - \epsilon)}}
 e^{  \frac{1}{2} 
     \int_0^t d t^\prime \int_0^t d t^{\prime \prime}
     v(t^\prime) B(t^\prime, t^{\prime \prime}) v(t^{\prime \prime})
     + \int_0^t d t^\prime v(t^\prime) C(t^\prime) + R(t)} 
\right|_{v=0},
\label{with_ABCR}
\end{eqnarray}
where the various kernels are
\begin{subequations}
\begin{align}
A(t^\prime, t^{\prime \prime}) 
&= i \delta (t^\prime - t^{\prime \prime}) 
&C(t^\prime)&=- (x - a) - i x t^\prime
\label{den_a}
\\
B(t^\prime, t^{\prime \prime})
&= - 
\left\{ 1 + i ~t_< (t^\prime, t^{\prime \prime}) \right\}
&R(t)&= 
-\frac{1}{2} (x - a)^2 
-\frac{i}{2} x^2 t.
\label{den_b}
\end{align}
\end{subequations}
Here  $t_< (t^\prime, t^{\prime \prime})$
stands for the function which
picks the minimum among $t^{\prime}$ and $t^{\prime \prime}$.
In this form, we notice that it is possible to carry out the operations
of taking the functional derivatives in Eq.~(\ref{with_ABCR})
because the exponents are quadratic in $v$.

To evaluate Eq.~(\ref{with_ABCR}), consider the discrete analog, constructed
from symmetric matrices $\bf A$ and $\bf B$, 
column vectors 
${\bf c}, {\bf x}$ and 
${\bf \nabla}= \frac{\partial}{\partial \mathbf{x}}$, 
and the number $r$, for which we have the identity
\begin{eqnarray}
e^{\frac{1}{2} {\bf \nabla^\text{T} \cdot A \cdot \nabla}}
e^{\frac{1}{2} {\bf x^\text{T} \cdot B \cdot x 
                 + c^\text{T} \cdot x} + r}
=
e^{\frac{1}{2} {\bf x^\text{T} \cdot B \cdot K \cdot x 
                    + c^\text{T} \cdot K \cdot x
                    + \frac{1}{2} c^\text{T} \cdot K \cdot A \cdot c} 
                    + r + \frac{1}{2} \text{Tr} \ln {\bf K}},
\label{mat_id}
\end{eqnarray}
where $\bf K$ is the solution to the matrix equation
\begin{equation}
\Big[{\bf 1 - A \cdot B} \Big] {\bf \cdot K} = {\bf 1}.
\label{1-ABK}
\end{equation}
Generalizing the identity (\ref{mat_id}) to functions,
and, using it to evaluate the functional
derivatives in Eq.~(\ref{with_ABCR}), and at the end setting $v=0$,
we get
\begin{eqnarray}
\psi (x,t) 
=
  \pi^{-\frac{1}{4}}
 ~e^{R(t) + \frac{1}{2} 
     \int_0^t d t^\prime 
     \int_0^t d t^{\prime \prime} 
     \int_0^t d t^{\prime \prime \prime} 
     C(t^\prime) K(t^\prime, t^{\prime \prime}) 
     A(t^{\prime \prime}, t^{\prime \prime \prime}) 
     C(t^{\prime \prime \prime}) 
     + \frac{1}{2} \text{Tr} \ln K},
\end{eqnarray}
where $K(t^\prime, t^{\prime \prime})$ is the solution 
to the integral equation [which is the meaning
of Eq.~(\ref{1-ABK}) for functions] 
\begin{equation}
\int_0^t d \tau^\prime
\left[
\delta(t^\prime - \tau^\prime)
- \int_0^t d \tau^{\prime \prime}
A(t^{\prime}, \tau^{\prime \prime})
B(\tau^{\prime \prime}, \tau^{\prime})
\right]
K(\tau^{\prime}, t^{\prime\prime})
= \delta (t^\prime - t^{\prime\prime}).
\label{1-ABK=1_fun}
\end{equation}
Using the expressions for 
$A(t^\prime, t^{\prime \prime})$ and 
$B(t^\prime, t^{\prime \prime})$
in Eqs.~(\ref{den_a}) and (\ref{den_b})
the above equations simplify to
\begin{eqnarray}
\psi (x,t) 
=  \pi^{-\frac{1}{4}}
 ~e^{R(t) + \frac{i}{2} 
     \int_0^t d t^\prime 
     \int_0^t d t^{\prime \prime} 
     C(t^\prime) K(t^\prime, t^{\prime \prime}) 
     C(t^{\prime \prime}) 
     + \frac{1}{2} \text{Tr} \ln K},
\label{psi_rckc}
\end{eqnarray}
where $K(t^\prime, t^{\prime \prime})$ 
and $\text{Tr} \ln K$ are to be determined 
using the integral equation
\begin{equation}
K(t^{\prime}, t^{\prime \prime})
+ i \int_0^t d \tau
\Big[ 1 + i t_< (t^{\prime}, \tau) \Big]
K(\tau, t^{\prime \prime})
= \delta (t^\prime - t^{\prime \prime}).
\label{intk}
\end{equation}

\subsection{Solution of integral equation}
We start by differentiating the above integral equation 
twice with respect to 
$t^\prime$ to obtain the differential equation satisfied by
$K(t^{\prime}, t^{\prime \prime})$:
\begin{equation}
\left[
\frac{\partial^2}{{\partial t^\prime}^2}
+ 1 \right]
K(t^{\prime}, t^{\prime \prime})
= \frac{\partial^2}{{\partial t^\prime}^2}
\delta (t^{\prime} - t^{\prime \prime}).
\end{equation}
The boundary conditions on $K(t^{\prime}, t^{\prime \prime})$
are read off from Eq.~(\ref{intk}) to be 
\begin{subequations}\label{bconk}
\begin{eqnarray}
K(0, t^{\prime \prime})
&=& - i  \int_0^t d \tau K(\tau, t^{\prime \prime})
\label{k(0)}, \\
K(t, t^{\prime \prime})
&=& 
K(0, t^{\prime \prime})
+  \int_0^t d \tau \,\tau K(\tau, t^{\prime \prime})
\label{k(t)}.
\end{eqnarray}
\end{subequations}
We have presumed that the delta function in 
Eq.~(\ref{intk}) does not contribute at 
$t^{\prime} = 0$ and $t^{\prime} = t$.
(But see below.)
In terms of a Green's function $M(t^{\prime}, t^{\prime \prime})$,
which satisfies 
\begin{equation}
\left[
\frac{\partial^2}{{\partial t^\prime}^2}
+ 1 \right]
M(t^{\prime}, t^{\prime \prime})
= \delta (t^{\prime} - t^{\prime \prime}),
\label{meqn}
\end{equation}
we can write
\begin{eqnarray}
K(t^{\prime}, t^{\prime \prime})
= \frac{\partial^2}{{\partial t^\prime}^2}
M(t^{\prime}, t^{\prime \prime})
= \delta (t^{\prime} - t^{\prime \prime})
-  M(t^{\prime}, t^{\prime \prime}).
\label{kandm}
\end{eqnarray}
The continuity conditions for
$M(t^{\prime}, t^{\prime \prime})$ are dictated by
the Green's function equation (\ref{meqn}) to be
\begin{subequations}\label{conteqn}
\begin{align}
\left\{ M(t^{\prime}, t^{\prime \prime})
\right\}_{t^\prime = t^{\prime \prime} + \delta}
- \left\{ M(t^{\prime}, t^{\prime \prime})
\right\}_{t^\prime = t^{\prime \prime} - \delta}
&=0,
\\
\left\{ 
M^\prime(t^{\prime}, t^{\prime \prime})
\right\}_{t^\prime = t^{\prime \prime} + \delta}
- \left\{ 
M^\prime(t^{\prime}, t^{\prime \prime})
\right\}_{t^\prime = t^{\prime \prime} - \delta}
&=1,
\end{align}
\end{subequations}
and the boundary conditions on $M(t^{\prime}, t^{\prime \prime})$
are prescribed by the boundary conditions on 
$K(t^{\prime}, t^{\prime \prime})$ in Eqs.~(\ref{k(0)}) 
and (\ref{k(t)}). 
We start by writing the solution to 
$M(t^{\prime}, t^{\prime \prime})$ in the form
\begin{equation}
M(t^{\prime}, t^{\prime \prime})
= \left\{
\begin{array}{ll}
\alpha (t^{\prime \prime}) \sin  t^\prime
+ \beta (t^{\prime \prime}) \cos  t^\prime,
& 0 \leq t^\prime < t^{\prime \prime} \leq t,
\\
\eta(t^{\prime \prime}) \sin  t^\prime
+ \xi (t^{\prime \prime}) \cos  t^\prime,
& 0 \leq t^{\prime \prime} < t^\prime \leq t,
\end{array}
\right.
\label{min4}
\end{equation}
in terms of four arbitrary constants.
Using the continuity conditions (\ref{conteqn}) to determine 
two of the four constants gives us
\begin{equation}
K(t^{\prime}, t^{\prime \prime})
= \delta (t^{\prime} - t^{\prime \prime}) 
- \alpha (t^{\prime \prime})   \sin  t^\prime
- \xi (t^{\prime \prime}) \cos  t^\prime
-  \sin  t_> \cos  t_<,
\label{kin2}
\end{equation}
where we have suppressed the $t^\prime$ and
$t^{\prime \prime}$ dependence in 
$t_<(t^\prime, t^{\prime \prime})$ and 
$t_>(t^\prime, t^{\prime \prime})$.
Using the above expression for $K(t^{\prime}, t^{\prime \prime})$
in Eqs.~(\ref{k(0)}) and (\ref{k(t)})
we get the equations determining 
$\alpha (t^{\prime \prime})$ and $\xi (t^{\prime \prime})$ 
to be
\begin{subequations}
\begin{eqnarray}
\alpha (t^{\prime \prime})
i  [ 1 - \cos  t ]
+ \xi (t^{\prime \prime})
[1 + i \sin  t]
&=&  i \cos  t \cos  t^{\prime \prime} 
- \sin t^{\prime \prime}, 
\\
\alpha (t^{\prime \prime})
 \cos  t 
- \xi (t^{\prime \prime})
\sin t
&=& - \cos t \cos t^{\prime \prime}, 
\end{eqnarray}
\end{subequations}
which easily yields
\begin{subequations}
\begin{eqnarray}
\alpha (t^{\prime \prime})
&=& - e^{-i  t} \cos  (t - t^{\prime \prime}),
\label{alpha}
\\
\xi (t^{\prime \prime})
&=& 
i e^{-i  (t-t'')} \cos t.
\label{xi}
\end{eqnarray}
\end{subequations}
Using Eqs.~(\ref{alpha}) and (\ref{xi}) in eqn. (\ref{kin2})
we can obtain the solution to 
$ K(t^{\prime}, t^{\prime \prime})$ in the form
\begin{equation}
K(t^{\prime}, t^{\prime \prime})
=\delta (t^{\prime} - t^{\prime \prime})
- i \cos  (t - t^\prime)
\cos  (t - t^{\prime \prime})
-   \sin  (t - t_<) \cos (t - t_>),\label{solutionK}
\end{equation}
which can be verified to satisfy the original integral 
equation (\ref{intk}) by substitution.

\subsection{Evaluation of $\text{Tr} \ln K$}\label{trlnk_app}

We start by observing that $K(t^\prime, t^{\prime \prime})$
and $M(t^\prime, t^{\prime \prime})$ are related by 
Eq.~(\ref{kandm}), and $M(t^\prime, t^{\prime \prime})$
satisfies Eq.~(\ref{meqn}). These observations guide us to
construct the eigenvalue equation 
\begin{equation}
\left[
\frac{\partial^2}{{\partial t^\prime}^2}
+ 1 \right]
\psi_n (t^{\prime})
= E_n \psi_n (t^{\prime})
\label{en}
\end{equation}
where we presume that there exists a Hilbert space 
where $\psi_n (t^{\prime})$ are orthonormal 
eigenfunctions satisfying 
the orthonormality condition
\begin{subequations}
\begin{align}
\int_0^t dt^{\prime}\, {\psi_n (t^{\prime})}^* \psi_m (t^{\prime}) 
= \delta_{nm}
\label{ortho}
\end{align}
and the completeness relation
\begin{align}
\sum_n \psi_n (t^{\prime}) \psi_n (t^{\prime \prime})^* 
= \delta (t^{\prime} - t^{\prime \prime}).
\label{comp}
\end{align}
\end{subequations}
We shall defer the derivation of the boundary conditions
on $\psi_n (t^{\prime})$. 

It is convenient to
evaluate the trace of an operator in terms of its
representation as a kernel. Let us denote
${\bf L} \equiv 
\frac{\partial^2}{{\partial t^\prime}^2} + 1$ 
and introduce the matrix representation
\begin{equation}
{\bf L} \equiv L(t^\prime, t^{\prime \prime})
= \sum_n E_n \psi_n (t^\prime) \psi_n (t^{\prime \prime})^*, 
\end{equation} 
in terms of which our original Eq.~(\ref{en}) 
takes the form
\begin{equation}
\int_0^t dt^{\prime \prime} L(t^\prime, t^{\prime \prime})
\psi_n (t^{\prime \prime})
= E_n \psi_n (t^{\prime}).
\end{equation}
In general, we can write the kernel associated with 
an arbitrary operator $F$ constructed out of ${\bf L}$
to be
\begin{equation}
F({\bf L}) \equiv F(t^\prime, t^{\prime \prime}) 
= \sum_n F(E_n) \psi_n (t^\prime) \psi_n (t^{\prime \prime})^*.
\end{equation}
As a particular example we can write the kernel
introduced in Eq.~(\ref{meqn}) to be
\begin{equation}
{\bf L}^{-1} = {\bf M} \equiv 
M(t^\prime, t^{\prime \prime}) 
= \sum_n \frac{1}{E_n}~ 
\psi_n (t^\prime) \psi_n (t^{\prime \prime})^*.
\label{linv}
\end{equation}
The trace of a kernel in this representation is well defined by
\begin{eqnarray}
\text{Tr}\,F({\bf L}) 
\equiv \int_0^t dt^\prime F(t^\prime, t^{\prime}) 
= \int_0^t dt^\prime \sum_n F(E_n) \psi_n (t^\prime) \psi_n (t^{\prime})^*
= \sum_n F(E_n),
\end{eqnarray}
where in the last step we have used the orthonormality relation 
(\ref{ortho}). 

Using the above prescription and using Eqs.~(\ref{kandm}) 
and (\ref{linv}) we can write the kernel representation
of the operator $K(t^{\prime}, t^{\prime \prime})$ to be
\begin{eqnarray}
K(t^{\prime}, t^{\prime \prime}) 
= \delta (t^{\prime} - t^{\prime \prime})
- \sum_n \frac{1}{E_n}
  \psi_n (t^\prime) \psi_n (t^{\prime \prime})^* 
\label{kpsipsi_a}
= \sum_n \left( 1 - \frac{1}{E_n} \right)
    \psi_n (t^\prime) \psi_n (t^{\prime \prime})^*,
\label{kpsipsi}
\end{eqnarray}
where in the last step we have used the 
completeness relation (\ref{comp}).
This implies that the eigenvalues of $\mathbf{K}$ are $(1-1/E_n)$.
Therefore, the trace of the logarithm 
of the kernel $K(t^{\prime}, t^{\prime \prime})$ 
is
\begin{eqnarray}
\text{Tr}\ln {\bf K} 
&=& \sum_n \ln \left[ 1 - \frac{1}{E_n} \right] 
\label{trke}.
\end{eqnarray}

Now we return to the derivation of the boundary
conditions on $\psi_n (t^\prime)$.  
We multiply $\psi_n (t^{\prime \prime})$ by
the integral equation for the kernel
$K(t^{\prime}, t^{\prime \prime})$ in Eq.~(\ref{intk}),
integrate with respect to $t^{\prime \prime}$,
and recognize that acting on $\psi_n$, $K$ may be replaced by
its eigenvalue:
\begin{equation}
\left( 1 - \frac{1}{E_n} \right) \psi_n (t^{\prime})
+ i \int_0^t d \tau
\Big[ 1 + i  t_< (t^{\prime}, \tau) \Big]
\left( 1 - \frac{1}{E_n} \right) \psi_n (\tau)
= \psi_n (t^{\prime}),
\end{equation}
which further simplifies into 
\begin{equation}
\psi_n (t^{\prime})
= - i  \left( 1 - E_n \right)
\int_0^t d \tau
\Big[ 1 + i  t_< (t^{\prime}, \tau) \Big] \psi_n (\tau).
\end{equation}
From the last equation we read off 
the boundary conditions on $\psi_n (t^{\prime})$ 
to be\footnote{To derive
these results from the boundary conditions on the kernels (\ref{bconk}),
we must include the $\delta$-function terms there: in effect
$$
K(0,t'')\to K(0,t'')-\delta(0-t'')=-M(0,t''),
\quad K(t,t'')\to K(t,t'')-\delta(t-t'')=-M(t,t'').
$$
Then the eigenfunction construction (\ref{kpsipsi})
correctly implies the boundary condition (\ref{psibc}).} 
\begin{subequations}\label{psibc}
\begin{eqnarray}
\psi_n (0) &=& - i 
\left( 1 - E_n \right)
\int_0^t d \tau \,\psi_n (\tau),
\label{bc1}
\\
\psi_n (t) &=& \psi_n (0) 
+\left( 1 - E_n \right)
\int_0^t d \tau \,\tau\, \psi_n (\tau)
\label{bc2}. 
\end{eqnarray}
\end{subequations}
Now in terms of $z_n=1-E_n$, the desired trace takes the form
\begin{equation}
\text{Tr}\,\ln {\bf K} 
= \sum_n \ln \left(1 - \frac{1}{z_n} \right)^{-1}=\ln\text{det}\, \mathbf{K} 
\quad\text{where}\quad
\text{det}\,{\bf K}
= \prod_n \left(1 - \frac{1}{z_n} \right)^{-1}
\label{trkl}.
\end{equation}

The solution to Eq.~(\ref{en}) is 
\begin{equation}
\psi_n (t') = A_n \cos \sqrt{z_n} t'
+ B_n \sin \sqrt{z_n} t',
\label{anbn}
\end{equation} 
where the coefficients
$A_n$ and $B_n$ are in principle determined 
by the boundary conditions. 
For the special case of $z_n =0$, the solution is
\begin{equation}
\psi_n (t') = c_1 t' + c_2
\end{equation}
which leads to the trivial solution 
$\psi_n (t') = 0$, because the only solution
to $c_1$ and $c_2$ allowed by the boundary 
conditions are $c_1 = 0$ and $c_2 = 0$.
Thus we conclude that $z_n = 0$ is not an eigenvalue.
Then we substitute the solutions 
(\ref{anbn}) into Eqs.~(\ref{bc1}) and (\ref{bc2})
to get the equations relating $A_n$ and $B_n$:
\begin{subequations}\label{ab_eqn}
\begin{eqnarray}
A_n \left[
1 + i \sqrt{z_n} \sin \sqrt{z_n} t
\right]
+ B_n i \sqrt{z_n}
\left( 1 - \cos \sqrt{z_n} t \right) 
&=& 0,
\\
A_n \sin \sqrt{z_n} t
- B_n \cos \sqrt{z_n} t 
&=& 0,
\end{eqnarray}
\end{subequations}
which leads to nontrivial solutions to $\psi_n (t')$
only when the eigenvalues satisfy the characteristic equation 
$P ({z}) = 0$, where the determinant of the coefficient matrix in
Eq.~(\ref{ab_eqn}) is
\begin{eqnarray}
P ({z}) \equiv
\cos \sqrt{z} t + i \sqrt{z} \sin \sqrt{z} t.
\label{pz_eqn}
\end{eqnarray}

As an illustrative example consider a polynomial
$Q^{(N)}(z)$ of order $N$, 
which can be written in terms of the roots $z_1, z_2, \ldots, z_N$ of 
the equation $Q^{(N)}(z) = 0$ as 
\begin{equation}
Q^{(N)}(z) = d_N (z - z_1) (z - z_2) \ldots (z - z_N).
\label{qn}
\end{equation}
In this form we can deduce
\begin{eqnarray}
\frac{Q^{(N)}(1)}{Q^{(N)}(0)}
=\left(1 - \frac{1}{z_1} \right)
\left(1 - \frac{1}{z_2} \right)
\ldots
\left(1 - \frac{1}{z_N} \right).
\end{eqnarray}
Presuming that the nonpolynomial $P (z)$ in Eq.~(\ref{pz_eqn}) has a 
similar infinite product representation 
in terms  of its roots, we can write 
\begin{equation}
\text{det} ~{\bf K}
= \prod_n \left(1 - \frac{1}{z_n} \right)^{-1}
=\frac{P (0)}{P (1)}= e^{-it}.
\label{detk1}
\end{equation}
Thus we have established the simple result
\begin{equation}
\mathrm{Tr}\,\ln\mathbf{K}=-it.\label{lndet}
\end{equation}

\section{Results and Conclusions}

Using the solution for the kernel (\ref{solutionK}) 
we can evaluate the first two terms in the exponent 
in Eq.~(\ref{psi_rckc}) to be 
\begin{equation}
R(t) + \frac{i}{2} 
     \tint_0^t d t^\prime 
     \tint_0^t d t^{\prime \prime} 
     C(t^\prime) K(t^\prime, t^{\prime \prime}) 
     C(t^{\prime \prime})
=
- \frac{1}{2} (x - a \cos t)^2
- \frac{i}{2} \left\{2a x \sin t
- \tfrac{1}{2} a^2 \sin 2 t \right\}.
\label{rckc}
\end{equation}
Then when we add $\frac12\text{Tr}\,\ln \mathbf{K}$ in 
Eq.~(\ref{lndet}) we obtain
\begin{equation}
\psi (x,t) = \pi^{-\frac{1}{4}}
 e^{- \frac{1}{2} (x - a \cos t)^2}
 e^{- \frac{i}{2} [t + 2 a x \sin t - \frac{1}{2} a^2 \sin 2 t]}
\end{equation}
which is the required solution (\ref{soughtresult}).

The reader might now rightfully object that we have honed a mighty
machine to crack open a peanut.  However, it seems to us that we have
performed a useful pedagogical purpose here in exposing explicitly
functional techniques which should have much broader applicability in nontrivial
contexts.  We will present such applications in a subsequent publication.
 

\begin{acknowledgments}
We are grateful to Shankar Sachithanandam, 
In\'es Cavero-Pel\'aez, Prachi Parashar, and Jeffrey Wagner
for very useful discussions.  We thank the US Department of Energy
for support of this research.  KAM is grateful to the Department of
Physics, Washington University, for hospitality and support.
\end{acknowledgments}




\begin{thebibliography}{99}
\bibitem{fradkin:1966}
E. S. Fradkin,
``Application of functional methods in quantum field theory
and quantum statistics (II),''
Nucl. Phys. {\bf 76}, 588-624 (1966), section 5.

\bibitem{fried}
To our knowledge, the term Fradkin's representation was first 
used by H. M. Fried in his book entitled 
{\it Basics of Functional Methods and Eikonal Models},
(Editions Fronti\`eres, France, 1990).

\bibitem{sch-z} J. Schwinger, 
``On the Green's functions of quantized fields. I - II,''
Proc. Natl. Acad. Sci. USA
{\bf 37} (7), 452-455, 455-459 (1951).

\bibitem{pnas}
For a recent historical introduction to Schwinger's ideas on
Green's functions, and the relation to Feynman's path integrals,
see S. S. Schweber, 
``The sources of Schwinger's Green's functions,''
Proc. Natl. Acad. Sci. USA {\bf 102} (22), 7783-7788 (2005).

\bibitem{schiff:qm}
See, for example,
L. I. Schiff,
{\it Quantum Mechanics}, 3rd ed.,
(McGraw-Hill Book Company, New York, 1955), 3rd ed., section 13.
\end{thebibliography}
\end{document}